\documentclass[twocolumn,english,aps,superscriptaddress,showpacs]{revtex4}

\usepackage{amssymb}
\usepackage{amsmath}
\usepackage{colordvi}
\usepackage[colorlinks]{hyperref}
\usepackage{amsthm}
\usepackage{subeqnarray}
\usepackage{cases}
\usepackage{enumerate}
\usepackage{graphicx}
\usepackage{epsfig}
\usepackage{epstopdf}
\usepackage{subfigure}
\usepackage{color}
\usepackage{hyperref}
\usepackage{url}
\usepackage{verbatim}

\def\be{\begin{equation}}
\def\ee{\end{equation}}
\def\bea{\begin{eqnarray}}
\def\eea{\end{eqnarray}}
\def\nn{\nonumber}

\begin{document}

\title{Cavity assisted single- and two-mode spin-squeezed states via
phase-locked atom-photon coupling}
\author{Yong-Chang Zhang}
\author{Xiang-Fa Zhou}
\email{xfzhou@ustc.edu.cn}
\author{Xingxiang Zhou}
\author{Guang-Can Guo}
\author{Zheng-Wei Zhou}
\email{zwzhou@ustc.edu.cn}
\affiliation{Key Laboratory of Quantum Information, Chinese Academy of Sciences, University of Science and Technology of China, Hefei, 230026, China}
\affiliation{Synergetic Innovation Center of Quantum Information
and Quantum Physics, University of Science and Technology of China, Hefei, 230026, China}

\begin{abstract}
We propose a scheme to realize the two-axis counter-twisting spin-squeezing
Hamiltonian inside an optical cavity with the aid of phase-locked atom-photon
coupling. By careful analysis and extensive simulation, we demonstrate that our
scheme is robust against dissipation caused by cavity loss and atomic
spontaneous emission, and it can achieve significantly higher squeezing than
one-axis twisting. We further show how our idea can be extended to generate
two-mode spin-squeezed states in two coupled cavities. Due to its easy
implementation and high tunability, our scheme is experimentally realizable with
current technologies.
\end{abstract}

\maketitle

\emph{Introduction. -- }
Since the early work of Kitagawa and Ueda\cite{Ueda} and others
\cite{JianMa,Wineland}, spin-squeezed states have attracted much interest due
to their close relations with quantum information processing
\cite{Inform1,Inform2,Helmerson,Micheli,Lee,Riedel} and precision metrology
\cite{Ueda,JianMa,Metro1,Metro2,Metro3}. In the original work of Kitagawa and Ueda\cite{Ueda},  
two mechanisms, namely one-axis twisting (OAT) and two-axis
counter-twisting (TACT), were proposed to generate spin-squeezed states.
Preparation of such novel states has been the subject of many studies in
various physical setups such as feedback systems\cite{Thomsen}, Bose-Einstein
condensates (BEC)
\cite{Riedel,Metro3,OATe1,OATe2,OATe3,TATt1,TATt2,TATt3,TATt4,Toroidal}, Rydberg
lattice clocks \cite{Gil}, and atomic systems in cavities
\cite{CavAtom1,CavAtom2,CavAtom3,CavAtom4,CavAtom5,Twocavities,CavAtom6}.
To the best of our knowledge, all experiments to date focus on OAT
spin-squeezing, whereas TACT spin-squeezed states have not been realized in
experiments yet.

In quantum metrology, it is theoretically demonstrated \cite{Ueda,JianMa} that
TACT states are fundamentally superior to OAT states because measurement systems
based on them can approach the Heisenberg limit in which the precision of the
measurement scales with $1/N$, $N$ being the number of particles in the system.
In contrast, the precision allowed by OAT states scales with $1/N^{2/3}$.
Hence, it remains a very important task to generate and exploit TACT
spin-squeezed states using methods and techniques within the reach of current
technologies. There have been a few theoretical proposals such as converting OAT
into effective TACT \cite{TATt1,TATt2,TATt3}, implementing TACT with molecular
states \cite{Helmerson,TATt4}, utilizing ultracold atoms in two cavities \cite{Twocavities},
employing feedbacks in the measurement system \cite{Thomsen}, and using toroidal BECs \cite{Toroidal}.
Nevertheless, due to the demanding experimental requirements of these schemes, it remains
experimentally challenging to generate TACT spin-squeezed states.

In this work, we propose a scheme to realize a TACT Hamiltonian in a
cavity-atom system. Our proposal relies on phase-locked coupling between atoms
and photons only. Since both the atoms and cavity modes are only
virtually excited, it has the important advantage of being largely immune to
atomic and cavity dissipation.
Further, our scheme can be easily generalized to generate two-mode spin-squeezed (TMSS)
states by coupling two cavities, which can be used to estimate two observables simultaneously even
when they do not commute. They are widely used in many quantum applications such
as entanglement demonstration \cite{TMSS1,TMSS3}, quantum teleportation
\cite{TMSSqt}, and quantum metrology \cite{TMSSqm}.
Considering the rapid advance in cavity technology including the availability
of high-finesse optical cavities and strong cavity-atom coupling
\cite{Cavity1,Cavity2,Cavity3,Cavity4,Cavity5,Cavity6,Cavity7}, our proposal can
be realized with no fundamental difficulty.

\begin{figure}[h]
\centering
  \includegraphics[width=0.5\columnwidth]{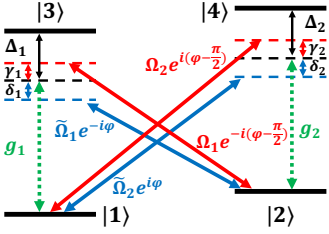}
\caption{(Color online) Atomic energy levels and transitions between them.
The complex Rabi frequencies $\widetilde{\Omega}_1 e^{-i\varphi}$,
$\widetilde{\Omega}_2 e^{i\varphi}, \Omega_1 e^{-i(\varphi-\frac{\pi}{2})}$, and
$\Omega_2 e^{i(\varphi-\frac{\pi}{2})}$ are associated with four phase-locked
driving lasers. $g_{1,2}$ is the coupling strength between the atom and the
cavity mode. $\Delta_{1,2}$, $\delta_{1,2}$ and $\gamma_{1,2}$ are detunings.}
\label{fig1}
\end{figure}

\emph{Effective Hamiltonian. -- }
We start by considering an ensemble of $N$ four-level atoms in an
optical cavity coupled to a single cavity mode and external laser fields. The
explicit level configuration is illustrated in Fig.\ref{fig1}, where $g_1$ and
$g_2$ are the cavity-atom coupling strengths driving the atomic transitions
$\vert 1\rangle \leftrightarrow \vert 3 \rangle$  and $\vert 2\rangle
\leftrightarrow \vert 4 \rangle$, $\widetilde{\Omega}_{1,2}$ and $\Omega_{1,2}$
are Rabi frequencies of the external laser fields, and $\Delta_{1,2}$,
$\delta_{1,2}$ and $\gamma_{1,2}$ are detunings. To realize the desired TACT
interaction, we also assume a fixed relative phase of $\pi/2$ ($-\pi/2$)
between $\Omega_1$($\Omega_2$) and
$\widetilde{\Omega}_1$($\widetilde{\Omega}_2$) \cite{Cho,Chen}.
The Hamiltonian reads
\bea
H &=&\sum^N_{j=1}\big\{\frac{e^{i\varphi}}{2}\left[\widetilde{\Omega}_2 e^{-i(\Delta_2 +\delta_2)t} - i \Omega_2 e^{-i(\Delta_2 -\gamma_2)t} \right] |1\rangle_j\langle 4|  \nn \\
&+& \frac{e^{-i\varphi}}{2}\left[\widetilde{\Omega}_1e^{-i(\Delta_1 +\delta_1)t} + i \Omega_1 e^{-i(\Delta_1 -\gamma_1)t} \right] | 2\rangle_j\langle 3|  \nn \\
&+& g_1| 1\rangle_j\langle 3| a^\dagger e^{-i\Delta_1 t} +g_2| 2\rangle_j\langle
4| a^\dagger e^{-i\Delta_2 t}  +h.c. \big\},
\label{eq1}
\eea
where $a^\dagger$($a$) is the creation (annihilation) operator
of the cavity mode, $\pm\varphi$ and $\pm(\varphi-\frac{\pi}{2})$ are the
phases of the external laser fields, and the detunings are defined as
$\Delta_{1(2)}=\omega_{3(4)}-\omega_{1(2)}-\omega_c$,
$\gamma_{1(2)}=\omega_{2(1)}-\omega_{1(2)}-\omega_c+\omega_{L_1(L_2)}$, and
$\delta_{1(2)}=\omega_{1(2)}-\omega_{2(1)}+\omega_c-\omega_{\widetilde{L}
_1(\widetilde{L}_2)}$ with $\omega_{L_{1,2},\widetilde{L}_{1,2}}$ and $\omega_c$
being the frequencies of the driving lasers and the cavity mode. The rotating
wave approximation was used to derive the Hamiltonian in Eq.(\ref{eq1}) in the
rotating frame defined by
$H_0=\sum^N_{j=1}\sum^4_{k=1}\omega_k| k\rangle_j\langle k| +\omega_c(a^\dagger
a+\frac{1}{2})$. To simplify our discussion, here and in the following, we
assume $\delta=\delta_1=\delta_2$, $\gamma=\gamma_1=\gamma_2$, and set
$\varphi=0$.
For large detunings with
\begin{equation}
| \Delta_{1,2}| , | \Delta_{1,2}+\delta| , | \Delta_{1,2}-\gamma| \gg | g_{1,2}| , | \Omega_{1,2} | , | \widetilde{\Omega}_{1,2} |,
\label{eq2}
\end{equation}
all the high energy levels can be adiabatically eliminated, leading to the
following effective Hamiltonian involving only the two lowest states and the
cavity mode,
\begin{eqnarray}
H'&=&\left\{c_z-c'_z\sin\big[ (\delta+\gamma)t\big ]\right \} S_z \nonumber \\
 &-&\big[ \frac{A}{2}S_xa^\dagger e^{i\delta t} +\frac{B}{2}S_ya^\dagger e^{-i\gamma t} +h.c.\big]. \label{eq3}
\end{eqnarray}
Here the collective atomic spin operators are defined as
$S_z=\frac{1}{2}\sum^N_{j=1}(| 1\rangle_j\langle 1| -| 2\rangle_j\langle 2|)$,
$S_x=\frac{1}{2}\sum^N_{j=1}(| 1\rangle_j\langle 2| +| 2\rangle_j\langle 1|)$,
and $S_y=\frac{i}{2}\sum^N_{j=1}(| 2\rangle_j\langle 1| -| 1\rangle_j\langle
2|)$. The explicit expressions for the coefficients $c_z$, $c_z'$, $A$, and $B$
can be found in the Supplementary Material \cite{SM}.

If we further assume that the effective couplings in Eq.(\ref{eq3}) are much
weaker than the detunings, i.e.,
\begin{equation}
|\delta|,\ |\gamma|,\ |\delta \pm \gamma| \gg N|A|/4,\ N|B|/4, \label{eq4}
\end{equation}
the cavity mode is virtually excited only and can be
adiabatically eliminated too. We then obtain the following effective
Hamiltonian
\begin{equation}
H_{eff}=c_zS_z-c_xS^2_x+c_yS^2_y \label{eq5}
\end{equation}
with $c_x=\frac{A^2}{4\delta}$ and $c_y=\frac{B^2}{4\gamma}$ \cite{SM}. This
is the celebrated Lipkin-Meshkov-Glick (LMG) model \cite{JianMa}.
When $c_z=0$ and $c_x=c_y=\chi$, it reduces to the standard TACT Hamiltonian in
\cite{Ueda}. Experimentally, all coefficients $c_{x,y,z}$ can be controlled by
adjusting the Rabi frequencies of the driving lasers. If necessary, $c_z$ can
also be compensated by an external magnetic field \cite{Duan}.

To characterize the degree of spin-squeezing, we introduce the parameter
\cite{Ueda,JianMa}
\begin{equation}
\xi^2=\frac{(\Delta S_\perp)^2_{min}}{S/2}.
\label{eq6}
\end{equation}
Here $S=N/2$ with ${\bf S}=(S_x,\ S_y,\ S_z)$ the total spin operator,
and $(\Delta S_\perp)^2_{min}=(\langle {\bf S}^2_\perp \rangle-\langle {\bf
S}_\perp \rangle^2)_{min}$ is the minimum spin fluctuation in the direction
perpendicular to the average spin $\langle {\bf S} \rangle$.
A state is a spin coherent state
(spin-squeezed state) if $\xi^2=1$ ($\xi^2<1$) \cite{Ueda}.

\emph{Numerical Simulation. -- }
In order to check the validity of our approximations, we numerically simulate the system evolution under the
effective Hamiltonian in Eq.(\ref{eq5}) and the original Hamiltonian in
Eq.(\ref{eq1}) and compare the results. To fulfill the approximations, the
explicit parameters are chosen as follows:
$g_{1,2}=\Omega_{1,2}=\widetilde{\Omega}_{1,2}=\Omega=5\times10^7s^{-1}$ ,
$\Delta_{1,2}=\Delta=10^9s^{-1}$, $\delta_{1,2}=10^8s^{-1}$,
$\gamma_{1,2}=1.26\times10^8s^{-1}$. With these parameters, the effective model
reduces to a standard TACT Hamiltonian with $c_z=0$ and
$\chi=5.69\times10^4s^{-1}$. We also assume that initially the cavity is in the
vacuum state and the atoms are all in the state $| 1 \rangle$, which
corresponds to a coherent spin state in the $z$ direction.
\begin{figure}[htb]
\centering
  \includegraphics[width=0.9\columnwidth]{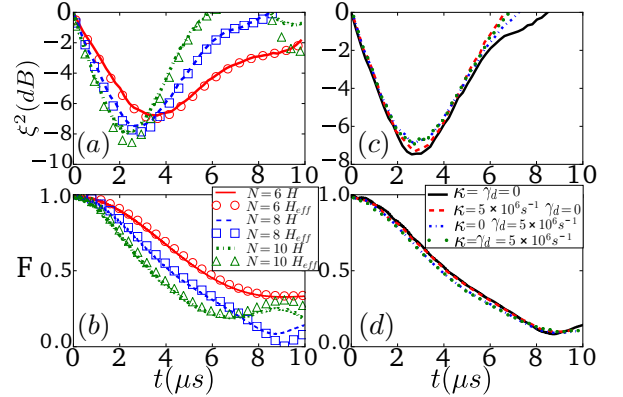}
\caption{(Color online) (a,b) Time evolution of $\xi^2$ and $F$ with
$H$ in Eq.(\ref{eq1}) and $H_{eff}$ in Eq.(\ref{eq5}) without dissipation for
$N=6$, $8$, and $10$.
  (c,d) Time evolution of $\xi^2$ and $F$ under the original Hamiltonian $H$ in
Eq.(\ref{eq1}) with dissipation for $N=8$. }
\label{fig2}
\end{figure}
Shown in Fig.\ref{fig2}(a,b) are the time dependences of the squeezing
parameter $\xi^2$ and the overlap function $F=|\langle \psi(0)|\psi(t)\rangle
|$ with the initial state of the system in the ideal case without cavity leakage
($\kappa=0$) and atomic spontaneous decay ($\gamma_d=0$). The state evolution
dictated by the effective TACT Hamiltonian in Eq.(\ref{eq5}) agrees very well
with that calculated directly from the original full Hamiltonian in
Eq.(\ref{eq1}), strong evidence that all approximations employed in our
derivation are reasonable.

\begin{figure}[htb]
\centering
  \includegraphics[width=0.49\columnwidth]{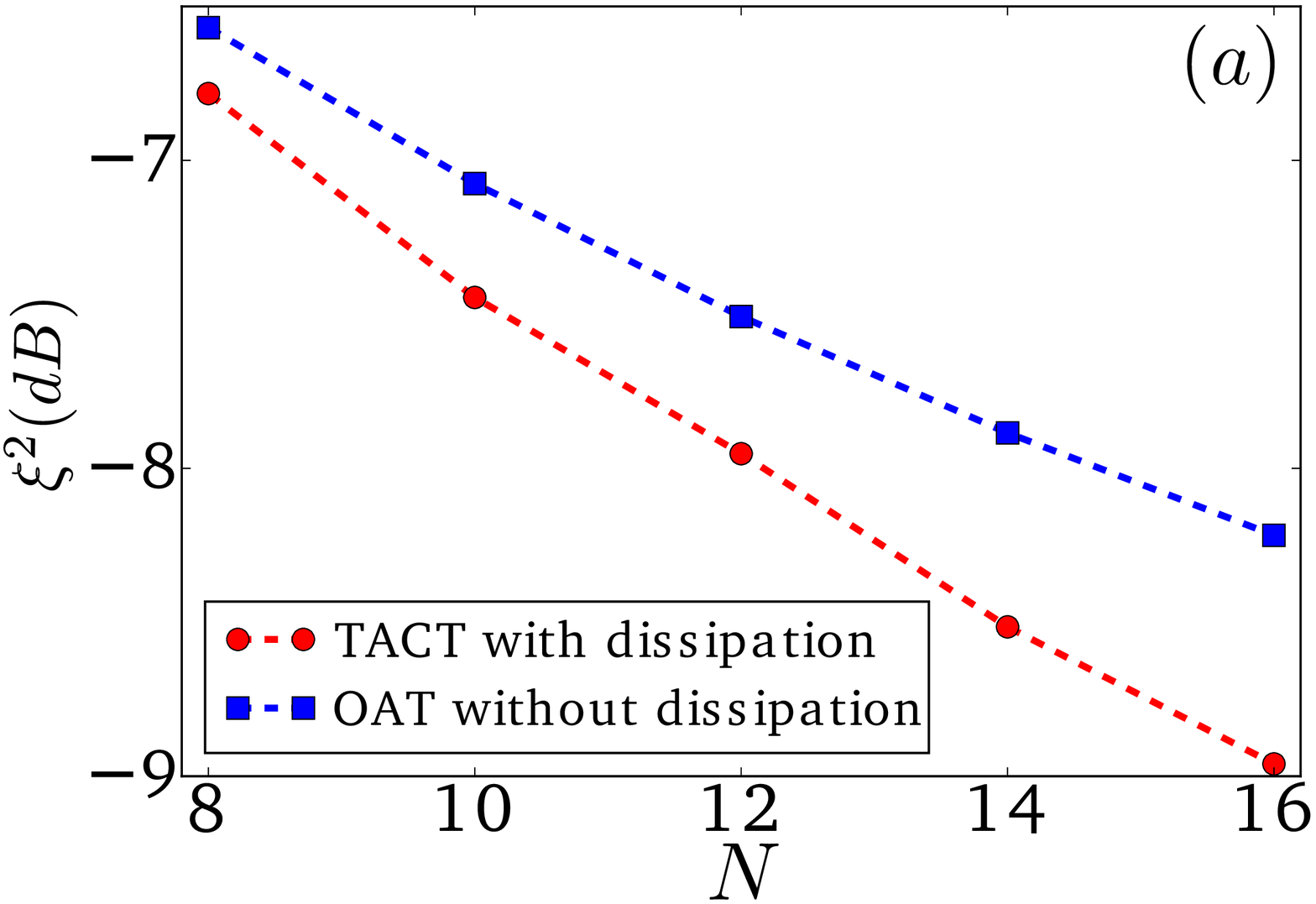}
  \includegraphics[width=0.49\columnwidth]{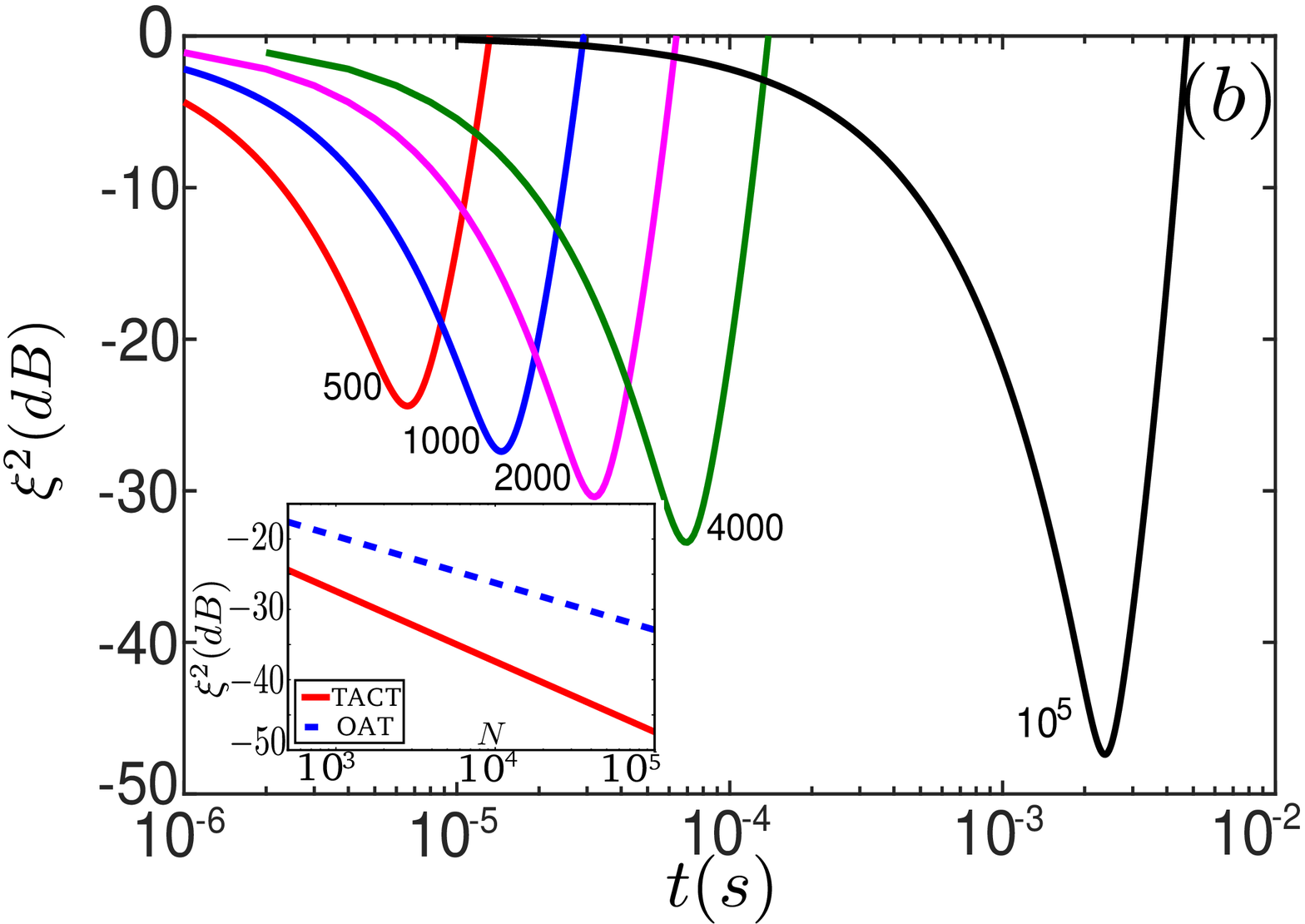}
\caption{(Color online) (a) Comparison of the maximum achievable squeezing in
our system with strong atomic and cavity dissipation rates
$\kappa=\gamma_d=5\times 10^6 s^{-1}$ and in a dissipation-free OAT
system $H_{OAT} = \chi S_x^2$. Other parameters are the same as
in Fig.\ref{fig2} except that $\Omega=2\times 10^7 s^{-1}$.
(b) Time evolution of $\xi^2$ in our system with $N=10^3-10^5$ atoms and no
dissipation. Shown in the inset is the maximum achievable squeezing in our
system and in OAT, both without dissipation.
Relevant parameters are $\Delta_1=-\Delta_2=1.88 \times
10^{10}s^{-1}$, $\gamma=2\delta=1.26 \times 10^{10}s^{-1}$ and
$A=B/\sqrt{2}=0.4\delta/N$ where $N=500,1000,2000,4000,$ and $10^5$. }
\label{fig3}
\end{figure}

We note in Fig.\ref{fig3}(b) that, though the maximum achievable squeezing
(i.e. the minimum $\xi^2$) increases with the number of atoms $N$ \cite{JianMa,TATt1}, the time it
takes to reach it increases with $N$ too. This is because
the nonlinear squeezing coefficients $c_x(c_y)$ in Eq.(\ref{eq5}) must decrease
with $N$ in order to maintain the virtual excitation of the system as dictated
by Eq.(\ref{eq4}). Though virtual excitation reduces the influence of the
dissipation, its eventual effect on squeezing must be carefully evaluated
because of the longer squeezing time required to reach the optimal squeezing.
For this purpose, we numerically
solve the master equation \cite{CavAtom4,Master1,Master2} of the system
\begin{equation}
\frac{\partial \rho}{\partial t}=-i[H,\ \rho]-\frac{\kappa}{2}\mathcal{D} (a,\rho)
-\frac{1}{2}\sum_{k=1}^N \sum_{s=1}^4 \gamma_{ks}\mathcal{D} (L_{ks},\rho).
\label{eq7}
\end{equation}
Here $\mathcal{D} (O,\rho)=O^\dagger O\rho+\rho
O^\dagger O-2O\rho O^\dagger$, $\rho$ is the density matrix, $\kappa$ and $\gamma_{ks}$ are the cavity
loss rate and atomic spontaneous decay rate, and $L_{k1}=|1\rangle_k \langle
3|$, $L_{k2}=|2\rangle_k \langle 3|$, $L_{k3}=|1\rangle_k \langle 4|$ and
$L_{k4}=|2\rangle_k \langle 4|$ are the jump operators.
The results are shown in Fig.\ref{fig2}(c,d) for N=8. It is seen that the
squeezing is robust against dissipation and the maximum
achievable squeezing is only slightly influenced by cavity loss and atomic
spontaneous emission as strong as $\kappa=\gamma_d=5\times10^6 s^{-1}$.
Since we have confirmed the validity of the virtual excitation of the cavity
mode in earlier simulations, we can adiabatically eliminate the cavity field
from the full Master equation to increase the scale of our simulated system.
This results in the following Master equation \cite{CavAtom4} that involves only the atomic
spin degrees of freedom,
\begin{equation}
  \begin{split}
\frac{\partial \rho_{eff}}{\partial t}&=-i[H_{eff},\ \rho_{eff}] -\frac{\gamma_d}{2}\sum_{\alpha=z,\pm}\sum^N_{k=1} a_\alpha \mathcal{D}(\sigma^k_\alpha,\rho_{eff}) \\
&-\frac{\kappa}{2}\left(\frac{A^2}{4\delta^2}\mathcal{D}(S_x,\rho_{eff})
+\frac{B^2}{4\gamma^2}\mathcal{D}(S_y,\rho_{eff}) \right),
\end{split}
  \label{eq:Master_S}
\end{equation}
where $\sigma^k_z=|1\rangle_k\langle1|-|2\rangle_k\langle2|$,
$\sigma^k_+=|1\rangle_k\langle2|$,
$\sigma^k_-=|2\rangle_k\langle1|$, and the explicit expressions for $a_{z,\pm}$ can
be found in the Supplementary Material \cite{SM}.
Using Eq.(\ref{eq:Master_S}), we can numerically simulate larger systems with
more atoms. In Fig.\ref{fig3}(a), we plot the maximum achievable
squeezing in our system with strong dissipation, as well as the maximum
squeezing attainable in an ideal OAT Hamiltonian with no dissipation. The
results show that, even in the presence of strong dissipation, our system can
achieve a higher degree of squeezing than what is possible with ideal OAT, and
the advantage grows with the size of the system. In Fig.\ref{fig3}(b), we
compare the maximum achievable squeezing of an ideal TACT Hamiltonian in
Eq.(\ref{eq5}) ($c_z=0$) with that of ideal OAT for larger system sizes on the
order of $10^3-10^5$. A large advantage is observed with our scheme. For a
system size of $N=10^5$ atoms as in recent experiments
\cite{CavAtom6,CavAtom7}, the ideal Hamiltonian Eq.(\ref{eq5}) for our system
can reach a squeezing of -47.4dB, significantly higher than current schemes
based on OAT \cite{OATe1,Metro2,Metro3,Riedel,CavAtom5} with
the same system size. Since the atomic decay time, estimated as
$1/\gamma_{eff}$ with $\gamma_{eff} \sim
\frac{\Omega^2}{4\Delta^2} \gamma_d \approx \frac{\delta \chi}{g^2}\gamma_d$
\cite{CavAtom1,CavAtom3,CavAtom5}, can be longer
than the time required to reach the optimal squeezing,
$t_o=1.58lnN/(3N\chi)$ \cite{TATt1}, e.g., when $N=10^5$, using parameters in
Fig.\ref{fig3}(b) with $g=1.26 \times 10^7 s^{-1}$ and $\gamma_d=3.77 \times 10^7 s^{-1}$, the
atomic decay time $1/\gamma_{eff}\approx 13ms$ is larger than
$t_o(10^5)\approx2.4ms$, and the influence of cavity
loss is much weaker than that of atoms' decay as illustrated in Fig.\ref{fig2}(c),
a high degree of squeezing can be achieved.

\emph{Two-mode spin-squeezed states. -- }
Our scheme can be extended to generate TMSS states \cite{TMSS1,TMSS3,TMSS2}
using two cavities. Assuming a coupling between the two cavity modes, we have
the following total Hamiltonian in the rotating frame,
\begin{equation}
H_{tc}=H_L+H_R-\widetilde{J}(a^\dagger_L a_R e^{i\Delta\omega t} +h.c.)
\label{eq9}
\end{equation}
where $a^\dagger_{L,R}(a_{L,R})$ is the creation (annihilation) operator for
the left and right cavity mode, $\widetilde{J}$ is the tunneling rate between
the cavities, $\Delta\omega=\omega^L_c-\omega^R_c$ is the detuning between the
two cavities with the local Hamiltonian $H_{\alpha \in
(L,R)}=-\frac{A_\alpha}{2}S^\alpha_x a^\dagger_\alpha e^{i\delta_\alpha t}
-\frac{B_\alpha}{2} S^\alpha_y a^\dagger_\alpha e^{-i\gamma_\alpha t} +h.c.$.
When the coefficients and detunings satisfy the following conditions
\begin{equation}
\begin{split}
&\delta_L=-\delta_R=\delta>0, \qquad -\gamma_L=\gamma_R=\gamma>0 \\
&\Delta\omega=\delta+\gamma, \quad A_L=A_R=A, \quad B_L=B_R=B
\end{split}
\label{eq10}
\end{equation}
the effective Hamiltonian for the two-cavity system can then be written as
\cite{SM}
\begin{equation}
H_{T}=\chi[(S^{L}_z)^2-(S^{R}_z)^2]+2J\chi(S^L_x S^R_y + S^L_y S^R_x)
\label{eq11}
\end{equation}
with $\chi=\frac{A^2}{4\delta}=\frac{B^2}{4\gamma}$, and
$J=\frac{\widetilde{J}}{\sqrt{\delta\gamma}}$. The second term in $H_{T}$ gives
rise to a TMSS state. The first term, which describes the on-site nonlinear
interaction in each cavity, has no contribution when $S^{L}_z=S^{R}_z$.

A TMSS state can be identified by checking that it satisfies the inequality
$\Delta'=(\Delta S^{(-)}_x)^2 + (\Delta S^{(+)}_y)^2 - \langle S^{(+)}_z\rangle
<0$ with $S^{(\pm)}_k=S^L_k \pm S^R_k$ ($k=x,y,z$) \cite{TMSS1,TMSS2,TMSS3}.
This criterion implies that fluctuations in nonlocal observables $S^-_x$ and
$S^+_y$ can be suppressed at the same time. Thus it is possible to achieve
higher measurement precisions for them simultaneously.
\begin{figure}[htbp]
\centering
  \includegraphics[width=0.9\columnwidth]{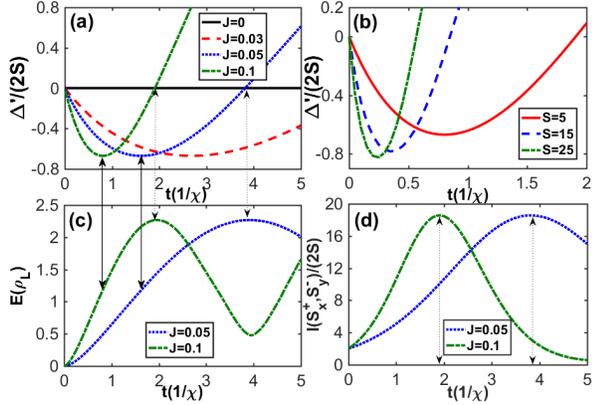}
\caption{(Color online) Time evolution of squeezing, entanglement, and quantum fisher
information of two-mode spin-squeezed states with the same total spin
number for the two modes. (a) $\Delta'$ vs $t$ while $S=N/2=5$, and $J=0,\,
0.03,\, 0.05,\, 0.1$; (b) $\Delta'$ vs $t$ while $J=0.1$, and $S=5,\, 15,\, 25$;
(c) Dependence of the von Neumann entropy $E(\rho_L)$ on time for
$J=0.05,\, 0.1$ when $S=5$; (d) The quantum Fisher information for
$|\Psi(t)\rangle$ at $J=0.05,\, 0.1$ when $S=5$.}
\label{fig4}
\end{figure}
When the total spins inside the two cavities are equal, $S_L=S_R$, a TMSS
state can be obtained by letting the system evolve under $H_T$ from an initial
state in which both cavities are in a coherent state: $|\Psi(0)\rangle=|S,S\rangle$ with $|m^L_i,m^R_j\rangle$
($m^{L,R}_i =-S,-S+1,\cdots,S-1,S$) the eigenvectors of $S_z$, and $S=N/2$ the
total spin. Plotted in Fig.\ref{fig4}(a,b) is the time evolution of
$\Delta'(t)$. One can see that $\Delta'$ is always zero when $J=0$ as both
cavities are decoupled in this case. When there is photon tunneling
between the cavities and thus $J\neq 0$, $\Delta'$ can become negative which
signals the emergence of TMSS states. Comparing Fig.\ref{fig4}(a) with
Fig.\ref{fig4}(b), we note that the time it takes to reach $\Delta'_{min}$, the
minimum value of $\Delta'$, is controlled by the coupling strength $J$,
and $\Delta'_{min}$ decreases as $S$ increases.
\begin{figure}[htbp]
\centering
  \includegraphics[width=0.55\columnwidth]{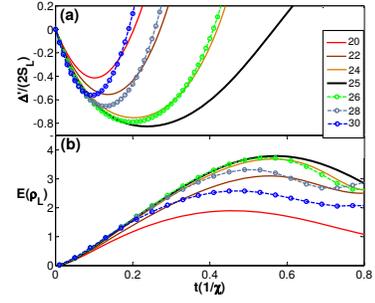}
\caption{(Color online) Time evolution of squeezing and entanglement of two-mode spin-
squeezed states with different total spin numbers for the two mode. (a) $\Delta'$ and (b) $E(\rho_L)$ vs $t$ for fixed $S_L=N_L/2=25$ and $J=0.1$.  $S_R$ varies from $20$ to $30$.}
\label{fig5}
\end{figure}
To investigate the entanglement of the TMSS state, we have further calculated
the von Neumann entropy $E(\rho_L)=-\rho_L\ln{\rho_L}$ of the reduced density
matrix $\rho_L=Tr_R(|\Psi\rangle\langle\Psi|)$, as well as the two-parameter
quantum Fisher information $I(S^+_x,S^-_y)_{ij}=2\langle \Psi|
\{H_i,H_j\}|\Psi\rangle-4\langle\Psi|H_i|\Psi\rangle\langle\Psi|H_j|\Psi\rangle$
\cite{Fisher} with $(i,j)=(1,2)$. Here, $H_{1(2)}=S^+_x(S^-_y)$, and $\{*,* \}$
is the anti-commutation relation. The results are shown in Fig.\ref{fig4}(c)
and (d). The TMSS state generated by the effective Hamiltonian (\ref{eq11})
leads to $I(S^+_x,S^-_y)_{11}=I(S^+_x,S^-_y)_{22}=I(S^+_x,S^-_y)$ [see
Fig.\ref{fig4}(d)] and $I(S^+_x,S^-_y)_{12,21}=0$. Comparing Fig.\ref{fig4}(c)
and (d) with Fig.\ref{fig4}(a), we find that $\Delta'$ reaches its minimum
(marked by black solid arrows) when $E(\rho_L)=E(\rho_L)_{max}/2$.
This result implies that the TMSS
state at $\Delta'_{min}$ is not a maximum entangled state. In addition, both
$I(S^+_x,S^-_y)$ and $E(\rho_L)$ attain their maxima only when $\Delta'$ evolves
back to zero.

To explore the influence of the imbalance between $S_L$ and $S_R$ on the TMSS
state, we fix $S_L$ and vary $S_R$ with the initial state $|S_L,S_R\rangle$. In
Fig.\ref{fig5}, the numerical result shows that $\Delta'_{min}$ reaches the
optimal value only when $S_L=S_R$, and increases as $\Delta S=|S_L-S_R|$
increases.  The time it takes to reach $\Delta'_{min}$, $t_o$, also decreases
with $\Delta S$. In contrast to the balanced case, $E(\rho_L)$ at $t_o$ is
smaller than $E(\rho_L)_{max}/2$, and it does not reach the maximum when
$\Delta'$ evolves back to zero, as shown in Fig.\ref{fig5}(b).  Therefore, to
obtain a TMSS state with lower $\Delta'$, it is helpful to prepare two cavities
with equal total spins.

\emph{Experimental Consideration. -- }
Experimentally, our model can be realized with an ensemble of $^{87}Rb$ atoms
in optical cavities \cite{CavAtom3,CavAtom4}. Two hyperfine states
$|F=1,m_F=1\rangle$ and $|F=2,m_F=2\rangle$ of the manifold $5S_{1/2}$ can be
used as the lower energy states $|1\rangle$ and $|2\rangle$ in Fig. \ref{fig1}.
Their energy splitting is $4.27 \times 10^{10}s^{-1}$. Two other hyperfine states of the manifold $5P_{1/2}$ with a
splitting of $5.03 \times 10^9 s^{-1}$ can be selected as the higher excited states $|3\rangle$ and
$|4\rangle$. This choice leads to a detuning of $\Delta_1-\Delta_2=3.77 \times
10^{10}s^{-1}$. To implement the effective TACT Hamiltonian in Eq.(\ref{eq5}) with
$c_z=0$ and $c_x=c_y$, we have a total of ten adjustable parameters, namely
$\Delta_{1,2}$, $\Omega_{1,2}$, $\widetilde{\Omega}_{1,2}$, $\delta_{1,2}$, and
$\gamma_{1,2}$. They need to satisfy the constraints
$\Delta_1-\Delta_2=3.77 \times 10^{10}s^{-1}$, $\frac{A^2}{\delta}=\frac{B^2}{\gamma}$,
and several others listed in the Supplementary Material \cite{SM}.
Since the number of these constraints is less than
the number of adjustable parameters, both the
TACT model and the LMG model can be achieved by adjusting the detunings and
couplings.

\emph{Conclusion. -- }
We have proposed a scheme to realize an effective TACT Hamiltonian in a
cavity-atom interacting system via phase-locked atom-photon coupling.
We proved that the approximations used in our derivation are justified and demonstrated that
greater degrees of squeezing can be achieved in our system than existing schemes
based on OAT. Furthermore, we generalized our ideas to a two-cavity system and showed how
TMSS states can be realized. Because of the high tunability of our scheme,
it is possible to access the full parameter ranges of the LMG model, enabling us to explore
its rich physics\cite{LMGQPT1,LMGQPT2,ESQPT1,ESQPT2,ESQPT3,ESQPT4}.

\emph{Acknowledgments. -- }
Y-C Zhang thanks Jun-Kang Chen for kind help on numerical simulation. This work was funded by the NKRDP (Grant No.2016YFA0301700), NNSFC (Grant Nos. 11574294, 61490711, 11474266), the Major Research plan of the NSFC (Grant No. 91536219), and the ``Strategic Priority Research Program(B)" of the Chinese Academy of Sciences (Grant No. XDB01030200).


\begin{thebibliography}{99}
\bibitem{Ueda}
Masahiro Kitagawa and Masahito Ueda, Phys. Rev. A \textbf{47}, 5138 (1993).

\bibitem{JianMa}
Jian Ma, Xiaoguang Wang, C. P. Sun and Franco Nori, Phys. Rep. \textbf{509}, 89 (2011).

\bibitem{Wineland}
 D.J. Wineland, J.J. Bollinger, W.M. Itano, F.L. Moore and D.J. Heinzen, Phys. Rev. A \textbf{46}, R6797 (1992).

\bibitem{Inform1}
Brian Julsgaard, Jacob Sherson, J. Ignacio Cirac, Jarom{\'i}r Fiur$\mathrm{\acute{a}\check{s}}$ek and Eugene S. Polzik, Nature \textbf{432}, 482 (2004).

\bibitem{Inform2}
Klemens Hammerer, Anders S. S$\mathrm{\o{}}$rensen and Eugene S. Polzik, Rev. Mod. Phys. \textbf{82}, 1041 (2010).

\bibitem{Helmerson}
Kristian Helmerson and Li You, Phys. Rev. Lett. \textbf{87}, 170402 (2001).

\bibitem{Micheli}
A. Micheli, D. Jaksch, J. I. Cirac and P. Zoller, Phys. Rev. A \textbf{67}, 013607 (2003).

\bibitem{Riedel}
Max F. Riedel, Pascal B\"{o}hi, Yun Li, Theodor W. H\"{a}nsch, Alice Sinatra and Philipp Treutlein, Nature \textbf{464}, 1170 (2010).

\bibitem{Lee}
Tony E. Lee, Florentin Reiter and Nimrod Moiseyev, Phys. Rev. Lett. \textbf{113}, 250401 (2014).

\bibitem{Metro1}
W. Muessel, H. Strobel, D. Linnemann, D. B. Hume and M. K. Oberthaler, Phys. Rev. Lett. \textbf{113}, 103004 (2014).

\bibitem{Metro2}
B. L\"{u}cke, M. Scherer, J. Kruse, L. Pezz$\mathrm{\acute{e}}$, F. Deuretzbacher, P. Hyllus, O. Topic, J. Peise, W. Ertmer, J. Arlt, L. Santos, A. Smerzi and C. Klempt, Science \textbf{334}, 773 (2011).

\bibitem{Metro3}
C. Gross, T. Zibold, E. Nicklas, J. Est$\mathrm{\grave{e}}$ve and M. K. Oberthaler, Nature \textbf{464}, 1165 (2010).

\bibitem{Thomsen}
L. K. Thomsen, S. Mancini and H. M. Wiseman, J. Phys. B: At. Mol. Opt. Phys. \textbf{35}, 4937 (2002).

\bibitem{OATe1}
J. Est$\mathrm{\grave{e}}$ve, C. Gross, A. Weller, S. Giovanazzi and M. K. Oberthaler, Nature \textbf{455}, 1216 (2008).

\bibitem{OATe2}
M. J. Martin, M. Bishof, M. D. Swallows, X. Zhang, C. Benko, J. von-Stecher, A. V. Gorshkov, A. M. Rey and Jun Ye, Science \textbf{341}, 632 (2013).

\bibitem{OATe3}
Jinling Lian, Lixian Yu, J.-Q. Liang, Gang Chen and Suotang Jia, Scientific Reports \textbf{3}, 3166 (2013).

\bibitem{TATt1}
Y. C. Liu, Z. F. Xu, G. R. Jin and L. You, Phys. Rev. Lett. \textbf{107}, 013601 (2011).

\bibitem{TATt2}
J. Y. Zhang, X. F. Zhou, G. C. Guo and Z. W. Zhou, Phys. Rev. A \textbf{90}, 013604 (2014).

\bibitem{TATt3}
Wen Huang, Yan-Lei Zhang, Chang-Ling Zou, Xu-Bo Zou and Guang-Can Guo, Phys. Rev. A \textbf{91}, 043642 (2015).

\bibitem{TATt4}
M. Zhang, Kristian Helmerson and L. You, Phys. Rev. A \textbf{68}, 043622 (2003).

\bibitem{Toroidal}
Tom$\mathrm{\acute{a}\check{s}}$ Opatrn$\mathrm{\acute{y}}$, Michal Kol$\mathrm{\acute{a}\check{r}}$ and Kunal K. Das, Phys. Rev. A \textbf{91}, 053612 (2015).

\bibitem{Gil}
L. I. R. Gil, R. Mukherjee, E. M. Bridge, M. P. A. Jones and T. Pohl, Phys. Rev. Lett. \textbf{112}, 103601 (2014).

\bibitem{CavAtom1}
F. Dimer, B. Estienne, A. S. Parkins and H. J. Carmichael, Phys. Rev. A \textbf{75}, 013804 (2007).

\bibitem{CavAtom2}
Anne E. B. Nielsen and Klaus M$\mathrm{\o{}}$lmer, Phys. Rev. A \textbf{77}, 063811 (2008).

\bibitem{CavAtom3}
Shi-Biao Zheng, Phys. Rev. A, \textbf{86}, 013828 (2012).

\bibitem{CavAtom4}
Emanuele G. Dalla Torre, Johannes Otterbach, Eugene Demler, Vladan Vuletic and Mikhail D. Lukin, Phys. Rev. Lett. \textbf{110}, 120402 (2013).

\bibitem{CavAtom5}
Lixian Yu, Jingtao Fan, Shiqun Zhu, Gang Chen, Suotang Jia and Franco Nori, Phys. Rev. A \textbf{89}, 023838 (2014).

\bibitem{Twocavities}
Caifeng Li, Jingtao Fan, Lixuan Yu, Gang Chen, Tian-Cai Zhang, and Suotang Jia, arXiv: quant-ph/1502.00470.

\bibitem{CavAtom6}
Monika H. Schleier-Smith, Ian D. Leroux, and Vladan Vuleti$\mathrm{\acute{c}}$, Phys. Rev. Lett. \textbf{104}, 073604 (2010).

\bibitem{CavAtom7}
Onur Hosten, Nils J. Engelsen, Rajiv Krishnakumar and Mark A. Kasevich, Nature \textbf{529}, 505 (2016).

\bibitem{Cavity1}
D. K. Armani, T. J. Kippenberg, S. M. Spillane and K. J. Vahala, Nature \textbf{421}, 925 (2003).

\bibitem{Cavity2}
S. M. Spillane, T. J. Kippenberg, K. J. Vahala, K. W. Goh, E. Wilcut and H. J. Kimble, Phys. Rev. A \textbf{71}, 013817 (2005).

\bibitem{Cavity3}
Takao Aoki, Barak Dayan, E. Wilcut, W. P. Bowen, A. S. Parkins, T. J. Kippenberg, K. J. Vahala and H. J. Kimble, Nature \textbf{443}, 671 (2006).

\bibitem{Cavity4}
Ferdinand Brennecke, Tobias Donner, Stephan Ritter, Thomas Bourdel, Michael K\"{o}hl and Tilman Esslinger, Nature \textbf{450}, 268 (2007).

\bibitem{Cavity5}
Yves Colombe, Tilo Steinmetz, Guilhem Dubois, Felix Linke, David Hunger and Jakob Reichel, Nature \textbf{450}, 272 (2007).

\bibitem{Cavity6}
Kater W. Murch, Kevin L. Moore, Subhadeep Gupta and Dan M. Stamper-Kurn, Nat. Phys. \textbf{4}, 561 (2008).

\bibitem{Cavity7}
Helmut Ritsch, Peter Domokos, Ferdinand Brennecke and Tilman Esslinger, Rev. Mod. Phys. \textbf{85}, 553 (2013).

\bibitem{TMSS1}
Brian Julsgaard, Alexander Kozhekin and Eugene S. Polzik, Nature, \textbf{413}, 400 (2001).

\bibitem{TMSS3}
D. W. Berry and B. C. Sanders, J. Phys. A: Math. Gen. \textbf{38}, L205 (2005).

\bibitem{TMSSqt}
D. W. Berry and B. C. Sanders, New J. Phys. \textbf{4}, 8 (2002).

\bibitem{TMSSqm}
C. Vaneph, T. Tufarelli and M. G. Genoni, Quant. Meas. Quant. Metrol. \textbf{1}, 12 (2013).

\bibitem{Cho}
Jaeyoon Cho, Dimitris G. Angelakis and Sougato Bose, Phys. Rev. Lett. \textbf{101}, 246809 (2008).

\bibitem{Chen}
Zhi-Xin Chen, Zheng-Wei Zhou, Xingxiang Zhou, Xiang-Fa Zhou, and Guang-Can Guo, Phys. Rev. A \textbf{81},
022303 (2010).

\bibitem{SM}
See supplementary material at URL for further details, which includes Refs. \cite{CavAtom4,James,tunneling1,tunneling2}.

\bibitem{James}
Daniel F.V. James and Jonathan Jerke, Can. J. Phys. \textbf{85}, 625 (2007).

\bibitem{tunneling1}
T. F. Roque and A. Vidiella-Barranco, J. Opt. Soc. Am. B \textbf{31}, 1232 (2014).

\bibitem{tunneling2}
Yong-Chun Liu, Yun-Feng Xiao, Xingsheng Luan, Qihuang Gong and Chee Wei Wong, Phys. Rev. A \textbf{91},033818 (2015).

\bibitem{Duan}
L.-M. Duan, E. Demler and M. D. Lukin, Phys. Rev. Lett. \textbf{91}, 090402 (2003).


\bibitem{Master1}
M. B. Plenio and P. L. Knight, Rev. Mod. Phys. \textbf{70}, 101 (1998).

\bibitem{Master2}
D. G. Norris, A. D. Cimmarusti, L. A. Orozco, P. Barberis-Blostein and H. J. Carmichael, Phys. Rev. A \textbf{86}, 053816 (2012).


\bibitem{TMSS2}
M. G. Raymer, A. C. Funk, B. C. Sanders and H. de Guise, Phys. Rev. A \textbf{67}, 052104 (2003).


\bibitem{Fisher}
Jing Liu, Xiao-Xing Jing and Xiaoguang Wang, Scientific Reports \textbf{5}, 8565 (2015).

\bibitem{LMGQPT1}
Octavio Casta$\mathrm{\tilde{n}}$os, Ram$\mathrm{\acute{o}}$n L$\mathrm{\acute{o}}$pez-Pe$\mathrm{\tilde{n}}$a, Jorge G. Hirsch and Enrique L$\mathrm{\acute{o}}$pez-Moreno, Phys. Rev. B \textbf{74}, 104118 (2006).

\bibitem{LMGQPT2}
F. de los Santos, E. Romera and O. Casta$\mathrm{\tilde{n}}$os, Phys. Rev. A \textbf{91}, 043409 (2015).

\bibitem{ESQPT1}
M. A. Caprio, P. Cejnar and F. Iachello, Ann. Phys. \textbf{323}, 1106 (2008).

\bibitem{ESQPT2}
Zi-Gang Yuan, Ping Zhang, Shu-Shen Li, Jian Jing and Ling-Bao Kong, Phys. Rev. A \textbf{85}, 044102 (2012).

\bibitem{ESQPT3}
G. Engelhardt, V. M. Bastidas, W. Kopylov and T. Brandes, Phys. Rev. A \textbf{91}, 013631 (2015).

\bibitem{ESQPT4}
P. Ribeiro, J. Vidal and R. Mosseri, Phys. Rev. E \textbf{78}, 021106 (2008).


\end{thebibliography}
\end{document}


\title{Supplementary material for ``Cavity assisted single- and two-mode spin squeezed states via phase-locked atom-photon coupling"}
\author{Yong-Chang Zhang}
\author{Xiang-Fa Zhou  \footnote{email: xfzhou@ustc.edu.cn}}
\author{Xingxiang Zhou}
\author{Guang-Can Guo}
\author{Zheng-Wei Zhou  \footnote{email: zwzhou@ustc.edu.cn}}
\affiliation{{Key Laboratory of Quantum Information, Chinese Academy of Sciences, University of Science and Technology of China, Hefei, 230026, China}\\
{Synergetic Innovation Center of Quantum Information and Quantum Physics, University of Science and Technology of China, Hefei, 230026, China}}

\maketitle

\setcounter{equation}{0}
\renewcommand{\theequation}{S\arabic{equation}}

\section{Single cavity: effective TACT Hamiltonian}

Assuming the parameters satisfy the following large detuning conditions,
$$\vert \Delta_{1,2}\vert , \vert \Delta_{1,2}+\delta\vert , \vert
\Delta_{1,2}-\gamma\vert \gg \vert g_{1,2}\vert , \vert \Omega_{1,2} \vert ,
\vert \widetilde{\Omega}_{1,2} \vert,$$ we can derive the following simplified
form for the system Hamiltonian in Eq.(1) in the main text by using the
effective Hamiltonian theory \cite{James}:

\allowdisplaybreaks

\begin{align}
\widetilde{H}&=-\frac{g^2_1}{\Delta_1} \big[ \sum^N_{j=1}\vert 1\rangle_j\langle 3\vert a^\dagger, \ \sum^N_{j=1}\vert 3\rangle_j\langle 1\vert a \big] -\frac{g^2_2}{\Delta_2} \big[ \sum^N_{j=1}\vert 2\rangle_j\langle 4\vert a^\dagger, \ \sum^N_{j=1}\vert 4\rangle_j\langle 2\vert a \big] \nonumber \\
&-\frac{1}{4}(\frac{\widetilde{\Omega}_1^2}{\Delta_1 +\delta_1} +\frac{\Omega_1^2}{\Delta_1 -\gamma_1}) \big[ \sum^N_{j=1}\vert 2\rangle_j\langle 3\vert, \ \sum^N_{j=1}\vert 3\rangle_j\langle 2\vert \big] -\frac{1}{4}(\frac{\widetilde{\Omega}_2^2}{\Delta_1 +\delta_2} +\frac{\Omega_2^2}{\Delta_2 -\gamma_2}) \big[ \sum^N_{j=1}\vert 1\rangle_j\langle 4\vert, \ \sum^N_{j=1}\vert 4\rangle_j\langle 1\vert \big] \nonumber \\
&-\frac{g_1 g_2}{2}(\frac{1}{\Delta_1} +\frac{1}{\Delta_2}) \big( \big[ \sum^N_{j=1}\vert 1\rangle_j\langle 3\vert a^\dagger, \ \sum^N_{j=1}\vert 4\rangle_j\langle 2\vert a \big]e^{-i(\Delta_1 -\Delta_2)t} +h.c.\big) \nonumber \\
&-\frac{i\widetilde{\Omega}_1 \Omega_1}{8}(\frac{1}{\Delta_1 +\delta_1} +\frac{1}{\Delta_1 -\gamma_1}) \big[ \sum^N_{j=1}\vert 2\rangle_j\langle 3\vert, \ \sum^N_{j=1}\vert 3\rangle_j\langle 2\vert \big] \big( e^{i(\delta_1 +\gamma_1)t} -e^{-i(\delta_1 +\gamma_1)t} \big) \label{eqA1} \\
&-\frac{i\widetilde{\Omega}_2 \Omega_2}{8}(\frac{1}{\Delta_2 +\delta_2} +\frac{1}{\Delta_2 -\gamma_2}) \big[ \sum^N_{j=1}\vert 1\rangle_j\langle 4\vert, \ \sum^N_{j=1}\vert 4\rangle_j\langle 1\vert \big] \big( e^{-i(\delta_2 +\gamma_2)t} -e^{i(\delta_2 +\gamma_2)t} \big) \nonumber \\
&-\frac{1}{4} \big\{ g_1\big( (\frac{\widetilde{\Omega}_1}{\Delta_1} +\frac{\widetilde{\Omega}_1}{\Delta_1 +\delta_1})e^{i\delta_1 t} -i(\frac{\Omega_1}{\Delta_1} +\frac{\Omega_1}{\Delta_1 -\gamma_1})e^{-i\gamma_1 t}\big) \big[ \sum^N_{j=1}\vert 1\rangle_j\langle 3\vert a^\dagger,\ e^{i\varphi}\sum^N_{j=1}\vert 3\rangle_j\langle 2\vert \big] \nonumber \\
&+g_2\big( (\frac{\widetilde{\Omega}_2}{\Delta_2} +\frac{\widetilde{\Omega}_2}{\Delta_1 +\delta_2}) e^{i\delta_2 t} +i(\frac{\Omega_2}{\Delta_2} +\frac{\Omega_2}{\Delta_2 -\gamma_2})e^{-i\gamma_2 t} \big) \big[ \sum^N_{j=1}\vert 2\rangle_j\langle 4\vert a^\dagger,\ e^{-i\varphi}\sum^N_{j=1}\vert 4\rangle_j\langle 1\vert \big] +h.c. \big\} \nonumber
\end{align}

Due to the large detunings, higher energy states are hardly excited. We can
then neglect all interaction terms involving $\vert 3\rangle$ and $\vert
4\rangle$. This leads to the following Hamiltonian,

\begin{align}
\widetilde{H}'=&-\frac{1}{4}(\frac{\widetilde{\Omega}_2^2}{\Delta_2 +\delta_2} +\frac{\Omega^2_2}{\Delta_2 -\gamma_2} +\frac{4 g^2_1}{\Delta_1}a^\dagger a)\sum^N_{j=1} \vert 1\rangle_j\langle 1\vert
-\frac{1}{4}(\frac{\widetilde{\Omega}_1^2}{\Delta_1 +\delta_1} +\frac{\Omega^2_1}{\Delta_1 -\gamma_1} +\frac{4 g^2_2}{\Delta_2}a^\dagger a)\sum^N_{j=1} \vert 2\rangle_j\langle 2\vert  \nonumber \\
&-\frac{i\widetilde{\Omega}_1 \Omega_1}{8}(\frac{1}{\Delta_1 +\delta_1} +\frac{1}{\Delta_1 -\gamma_1})\sum^N_{j=1}\vert 2\rangle_j\langle 2\vert \big( e^{i(\delta_1 +\gamma_1)t} -e^{-i(\delta_1 +\gamma_1)t} \big) \nonumber \\
&-\frac{i\widetilde{\Omega}_2 \Omega_2}{8}(\frac{1}{\Delta_2 +\delta_2} +\frac{1}{\Delta_2 -\gamma_2})\sum^N_{j=1}\vert 1\rangle_j\langle 1\vert \big( e^{-i(\delta_2 +\gamma_2)t} -e^{i(\delta_2 +\gamma_2)t} \big) \label{eqA2} \\
&-\frac{1}{4} \big\{ \big( g_1\widetilde{\Omega}_1 e^{i\varphi} (\frac{1}{\Delta_1} +\frac{1}{\Delta_1 +\delta_1})\sum^N_{j=1}\vert 1\rangle_j\langle 2\vert e^{i\delta_1 t} +g_2\widetilde{\Omega}_2 e^{-i\varphi} (\frac{1}{\Delta_2} +\frac{1}{\Delta_2 +\delta_2})\sum^N_{j=1}\vert 2\rangle_j\langle 1\vert e^{i\delta_2 t} \big)a^\dagger \nonumber \\
&+i\big( -g_1\Omega_1 e^{i\varphi} (\frac{1}{\Delta_1} +\frac{1}{\Delta_1 -\gamma_1})\sum^N_{j=1}\vert 1\rangle_j\langle 2\vert e^{-i\gamma_1 t} +g_2\Omega_2 e^{-i\varphi} (\frac{1}{\Delta_2} +\frac{1}{\Delta_2 -\gamma_2})\sum^N_{j=1}\vert 2\rangle_j\langle 1\vert e^{-i\gamma_2 t}\big) a^\dagger +h.c.\big\} \nonumber
\end{align}
Under the following assumptions for the convenience of presentation,
\begin{equation}
\begin{split}
&\delta=\delta_1=\delta_2, \qquad \gamma=\gamma_1=\gamma_2, \\
&A=g_1\widetilde{\Omega}_1(\frac{1}{\Delta_1} +\frac{1}{\Delta_1
+\delta_1})=g_2\widetilde{\Omega}_2(\frac{1}{\Delta_2} +\frac{1}{\Delta_2
+\delta_2}), \\
&B=g_1\Omega_1(\frac{1}{\Delta_1} +\frac{1}{\Delta_1
-\gamma_1})=g_2\Omega_2(\frac{1}{\Delta_2} +\frac{1}{\Delta_2 -\gamma_2}),
\end{split}
\label{eqA3}
\end{equation}
the Hamiltonian (\ref{eqA2}) is further simplified to

\begin{equation}
\widetilde{H}'=\big[c_z-c'_z\sin{\big( (\delta+\gamma)t\big)}\big]S_z-\big[
\frac{A}{2}(S_x\cos{\varphi}-S_y\sin{\varphi})a^\dagger e^{i\delta' t}
+\frac{B}{2}(S_x\sin{\varphi}+S_y\cos{\varphi})a^\dagger e^{-i\gamma' t}
+h.c.\big]
\label{eqA4}
\end{equation}
with
$\delta'=\delta-\frac{N}{2}(\frac{g^2_1}{\Delta_1}+\frac{g^2_2}{\Delta_2})$,
$\gamma'=\gamma+\frac{N}{2}(\frac{g^2_1}{\Delta_1}+\frac{g^2_2}{\Delta_2})$,
$c_z=\frac{1}{4}(\frac{\widetilde{\Omega}_1^2}{\Delta_1 +\delta_1}
+\frac{\Omega^2_1}{\Delta_1 -\gamma_1}-\frac{\widetilde{\Omega}_2^2}{\Delta_2
+\delta_2} -\frac{\Omega^2_2}{\Delta_2
-\gamma_2})-\frac{1}{2}(\frac{g^2_1}{\Delta_1}-\frac{g^2_2}{\Delta_2})a^\dagger
a$, $c'_z=\frac{1}{4} (\frac{\widetilde{\Omega}_1 \Omega_1}{\Delta_1 +\delta_1}
+\frac{\widetilde{\Omega}_1 \Omega_1}{\Delta_1 -\gamma_1}
+\frac{\widetilde{\Omega}_2 \Omega_2}{\Delta_2 +\delta_2}
+\frac{\widetilde{\Omega}_2 \Omega_2}{\Delta_2 -\gamma_2})$,
$S_z=\frac{1}{2}\sum^N_{j=1}(\vert 1\rangle_j\langle 1\vert -\vert
2\rangle_j\langle 2\vert)$, $S_x=\frac{1}{2}\sum^N_{j=1}(\vert 1\rangle_j\langle
2\vert +\vert 2\rangle_j\langle 1\vert)$, and $S_y=\frac{i}{2}\sum^N_{j=1}(\vert
2\rangle_j\langle 1\vert -\vert 1\rangle_j\langle 2\vert)$. In deriving
(\ref{eqA4}), we have neglected a constant term $\frac{1}{8}\big[
(\frac{\widetilde{\Omega}_1^2}{\Delta_1 +\delta_1} +\frac{\Omega^2_1}{\Delta_1
+\gamma_1}+\frac{\widetilde{\Omega}_2^2}{\Delta_2 +\delta_2}
+\frac{\Omega^2_2}{\Delta_2 -\gamma_2}) +(\frac{\widetilde{\Omega}_1
\Omega_1}{\Delta_1 +\delta_1} +\frac{\widetilde{\Omega}_1 \Omega_1}{\Delta_1
-\gamma_1} -\frac{\widetilde{\Omega}_2 \Omega_2}{\Delta_2 +\delta_2}
-\frac{\widetilde{\Omega}_2 \Omega_2}{\Delta_2 -\gamma_2})\sin{(
(\delta+\gamma)t)} \big]\sum^N_{j=1}(\vert 1\rangle_j\langle 1\vert +\vert
2\rangle_j\langle 2\vert)$. Initially, the cavity mode is in vacuum state, and
it can only be virtually excited due to the large detuning condition. By
adiabatically eliminating the cavity mode, we have the following effective
Hamiltonian,

\begin{equation}
H_{eff}=c_zS_z-c_x(S_x\cos{\varphi}-S_y\sin{\varphi})^2+c_y(S_x\sin{\varphi}+S_y\cos{\varphi})^2
\label{eqA5}
\end{equation}
with $c_x=\frac{A^2}{4\delta'}$ and $c_y=\frac{B^2}{4\gamma'}$. For
simplicity, we assume the following optional conditions in the main text and
the rest of this Supplementary material: $\delta'\approx \delta$,
$\gamma'\approx \gamma$, and $\delta,\gamma
\gg\frac{N}{2}(\frac{g^2_1}{\Delta_1}+\frac{g^2_2}{\Delta_2})$.
If we fix $\varphi=0$, the effective Hamiltonian in Eq.(\ref{eqA5}) then
becomes Eq.(5) in the main text.


Similarly, by adiabatically eliminating the atomic excitation and cavity
modes from the full master equation Eq.(7) \cite{CavAtom4} in the main text, an
effective master equation involving only the spin degree of freedom can be
derived as in the following
\begin{equation}
\begin{split}
\frac{\partial \rho_{eff}}{\partial t}=&-i[H_{eff},\ \rho_{eff}]-\frac{\kappa}{2}\big[\frac{A^2}{4\delta^2}(S^2_x \rho_{eff} +\rho_{eff} S^2_x -2S_x \rho_{eff} S_x) \\
&\qquad+\frac{B^2}{4\gamma^2}(S^2_y \rho_{eff} +\rho_{eff} S^2_y -2S_y \rho_{eff} S_y) \big]\\
&-\frac{\gamma_d}{2}\big[ a_z\sum^N_{k=1}\left((\sigma^k_z)^2\rho_{eff}+\rho_{eff}(\sigma^k_z)^2-2\sigma^k_z\rho_{eff}\sigma^k_z\right)\\
&\qquad +a_-\sum^N_{k=1}\left(\sigma^k_+\sigma^k_-\rho_{eff}+\rho_{eff}\sigma^k_+\sigma^k_- -2\sigma^k_-\rho_{eff}\sigma^k_+\right) \\
&\qquad+a_+\sum^N_{k=1}\left(\sigma^k_-\sigma^k_+\rho_{eff}+\rho_{eff}
\sigma^k_-\sigma^k_+ -2\sigma^k_+\rho_{eff}\sigma^k_-\right)\big],
\end{split}
\label{eq1}
\end{equation}
where $\sigma^k_z=|1\rangle_k\langle1|-|2\rangle_k\langle2|$, $\sigma^k_+=|1\rangle_k\langle2|$, $\sigma^k_-=|2\rangle_k\langle1|$, and
\begin{equation}
\begin{split}
&a_z=\frac{1}{4}\left(\frac{\Omega^2_1}{(\Delta_1-\gamma)^2} +\frac{\widetilde{\Omega}^2_1}{(\Delta_1+\delta)^2} +\frac{\Omega^2_2}{(\Delta_2-\gamma)^2} +\frac{\widetilde{\Omega}^2_2}{(\Delta_2+\delta)^2} \right)\\
&a_-=\frac{1}{4}\left(\frac{\Omega^2_2}{(\Delta_2-\gamma)^2} +\frac{\widetilde{\Omega}^2_2}{(\Delta_2+\delta)^2} \right) \\
&a_+=\frac{1}{4}\left(\frac{\Omega^2_1}{(\Delta_1-\gamma)^2} +\frac{\widetilde{\Omega}^2_1}{(\Delta_1+\delta)^2} \right)
\end{split}
\label{eq2}
\end{equation}

\section{Two cavity: effective TMSS Hamiltonian}
The two cavities are coupled by inter-cavity photon tunneling
\cite{tunneling1,tunneling2}.
In the interaction picture, the total Hamiltonian reads

\begin{equation}
\begin{split}
H_{tc}=&-\frac{A_L}{2}S^L_x a^\dagger_L e^{i\delta_L t} -\frac{B_L}{2}S^L_y a^\dagger_L e^{-i\gamma_L t}\\
&-\frac{A_R}{2}S^R_x a^\dagger_R e^{i\delta_R t} -\frac{B_R}{2}S^R_y a^\dagger_R e^{-i\gamma_R t} \\
&-\widetilde{J}a^\dagger_L a_R e^{i\Delta \omega t} +h.c.,
\end{split}
\label{eqA6}
\end{equation}
where $\Delta\omega=\omega^L_c-\omega^R_c$ is the frequency detuning between
the two cavities. Unlike in Eq.(\ref{eqA4}), here we have fixed
$\varphi=0$ and neglected the Zeeman term $S^\alpha_z$ as it does not
contribute to the effective coupling between the two spin systems and can be
compensated by external fields. If the coefficients and detunings satisfy the
relations
\begin{equation}
\begin{split}
&\delta_L=-\delta_R=\delta>0, \qquad -\gamma_L=\gamma_R=\gamma>0 \\
&\Delta\omega=\delta+\gamma, \quad A_L=A_R=A, \quad B_L=B_R=B
\end{split}
\end{equation}
we can derive the following effective Hamiltonian by keeping terms to
third order in the effective Hamiltonian \cite{James},

\begin{equation}
\begin{split}
H_{TMSS}=&-\frac{A^2}{4\delta}(S^L_x)^2-\frac{B^2}{4\gamma}(S^L_y)^2+\frac{A^2}{4\delta}(S^R_x)^2+\frac{B^2}{4\gamma}(S^R_y)^2+\frac{\widetilde{J}AB}{2\delta \gamma}(S^L_x S^R_y + S^L_y S^R_x)\\
=&\chi[(S^L_z)^2-(S^L)^2]-\chi[(S^R_z)^2-(S^R)^2] +2J\chi (S^L_x S^R_y + S^L_y S^R_x)
\end{split}
\label{eqA7}
\end{equation}
with $\chi=\frac{A^2}{4\delta}=\frac{B^2}{4\gamma}$, and
$J=\frac{\widetilde{J}}{\sqrt{\delta\gamma}}$. Here, we have used the relation
$(S^\alpha_x)^2 +(S^\alpha_y)^2 +(S^\alpha_z)^2=(S^\alpha)^2$ ($\alpha=L,R$).
The constant term, $\chi[(S^R)^2-(S^L)^2]$, can be neglected. It should be
emphasized that the signs of the coefficients of $(S^{L,R}_z)^2$ must be
opposite in order to generate TMSS states.